\documentclass{emulateapj}

\usepackage{psfig}
\usepackage{amsmath}
\usepackage{amssymb}
\usepackage{graphicx}
\usepackage{appendix}
\usepackage{color}
\usepackage{blindtext}
\setcitestyle{citesep={,}}

\newcommand{\lSect}[1]{{\label{sec:#1}}}
\newcommand{\lFig}[1]{{\label{fig:#1}}}
\newcommand{\lEq}[1]{{\label{eq:#1}}}

\def\gtaprx {\lower .1ex\hbox{\rlap{\raise .6ex\hbox{\hskip .3ex
	{\ifmmode{\scriptscriptstyle >}\else
		{$\scriptscriptstyle >$}\fi}}}
	\kern -.4ex{\ifmmode{\scriptscriptstyle \sim}\else
		{$\scriptscriptstyle\sim$}\fi}}}
\def\ltaprx {\lower .1ex\hbox{\rlap{\raise .6ex\hbox{\hskip .3ex
	{\ifmmode{\scriptscriptstyle <}\else
		{$\scriptscriptstyle <$}\fi}}}
	\kern -.4ex{\ifmmode{\scriptscriptstyle \sim}\else
		{$\scriptscriptstyle\sim$}\fi}}}
\newcommand{\FIGFF}[2]{{\ref{fig:#2}{#1}}}

\newcommand{\FIG}[2]{{Fig.~\FIGFF{#1}{#2}}}
\newcommand{\Fig}[1]{{\FIG{}{#1}}}

\newcommand{\Sectff}[1]{{\ref{sec:#1}}}
\newcommand{\Sect}[1]{{\S\Sectff{#1}}}

\newcommand{\Eqref}[1]{{\ref{eq:#1}}}
\newcommand{\Eqff}[1]{{(\Eqref{#1})}}

\newcommand{\Eq}[1]{{eq.~\Eqff{#1}}}

\newcommand{\Leg}[1]{{\textit{#1}}}
\newcommand{\Msun}{\ensuremath{\mathrm{M}_\odot}}

\newcommand{\SEDONA}{\ensuremath{\mathrm{\texttt{SEDONA}}}}

\def\gtaprx {\lower .1ex\hbox{\rlap{\raise .6ex\hbox{\hskip .3ex
	{\ifmmode{\scriptscriptstyle >}\else
		{$\scriptscriptstyle >$}\fi}}}
	\kern -.4ex{\ifmmode{\scriptscriptstyle \sim}\else
		{$\scriptscriptstyle\sim$}\fi}}}
\def\ltaprx {\lower .1ex\hbox{\rlap{\raise .6ex\hbox{\hskip .3ex
	{\ifmmode{\scriptscriptstyle <}\else
		{$\scriptscriptstyle <$}\fi}}}
	\kern -.4ex{\ifmmode{\scriptscriptstyle \sim}\else
		{$\scriptscriptstyle\sim$}\fi}}}

\begin{document}

\submitted{8 May, 2018}
\accepted{22 Mar, 2019}
\shortauthors{Tuguldur Sukhbold}

\title{Properties of Type-I\MakeLowercase{a} Supernova Light Curves}

\author{Tuguldur Sukhbold\altaffilmark{1,2,3}}
\altaffiltext{1}{NASA Hubble Fellow, e-mail: tuguldur.s@gmail.com}
\altaffiltext{2}{Department of Astronomy, The Ohio State University, Columbus, OH 43210, USA}
\altaffiltext{3}{Center for Cosmology and AstroParticle Physics, The Ohio State University, Columbus, OH 43210, USA}

\begin{abstract}
I show that the characteristic diffusion timescale and the gamma-ray escape timescale, of SN Ia supernova ejecta, are related with each other through the time when the bolometric luminosity, $L_{\rm bol}$, intersects with instantaneous radioactive decay luminosity, $L_\gamma$, for the second time after the light-curve peak. Analytical arguments, numerical radiation-transport calculations, and observational tests show that $L_{\rm bol}$ generally intersects $L_\gamma$ at roughly $1.7$ times the characteristic diffusion timescale of the ejecta. This relation implies that the gamma-ray escape timescale is typically 2.7 times the diffusion timescale, and also implies that the bolometric luminosity 15 days after the peak, $L_{\rm bol}(t_{15})$,  must be close to the instantaneous decay luminosity at that time, $L_\gamma (t_{15})$. With the employed calculations and observations, the accuracy of $L_{\rm bol}=L_\gamma$ at $t=t_{15}$ is found to be comparable to the simple version of ``Arnett's rule'' ($L_{\rm bol}=L_\gamma$ at $t=t_{\rm peak}$). This relation aids the interpretation of SN Ia supernova light curves and may also be applicable to general hydrogen-free explosion scenarios powered by other central engines.
\end{abstract}


\section{INTRODUCTION}
\lSect{intro}

Supernovae of Type-Ia are believed to result from thermonuclear explosions of white dwarfs \citep{Hoy60}. While they play a major role as a cosmographic tool, the identity of the progenitors and the nature of the ignition process remain a mystery \citep[for a general review, see][]{Mao14}. In this work, however, an agnostic stance is taken on the exact nature of the progenitor or explosion, and instead I focus on the generic properties of the light curves.

It is well known that the optical display of a Type-Ia explosion is predominantly powered by the radioactive decay chain of $^{56}$Ni $\rightarrow ^{56}$Co $\rightarrow ^{56}$Fe \citep{Pan62,Tru67,Bod68,Col69},\footnote{Though he did not provide a correct description of Si-burning, Dr. Titus Pankey Jr. appears to have been the first to speculate on the connection between the decay of $^{56}$Ni and Type-Ia light curves.} whose power is a precisely known exponentially decaying function of time \citep[e.g.,][]{Nad94}, 
\begin{equation}
L_\gamma = \frac{M_{\rm Ni}}{\Msun}(C_{\rm Ni}e^{-t/\tau_{\rm Ni}}+C_{\rm Co}e^{-t/\tau_{\rm Co}})\ \rm ergs\ s^{-1},
\end{equation}
where $t$ and $M_{\rm Ni}$ are time and $^{56}$Ni mass, and $C_{\rm Ni}\approx6.45\times10^{43}$, $C_{\rm Co}\approx1.45\times10^{43}$, $\tau_{\rm Ni}=8.8$ days and $\tau_{\rm Co}=111.3$ days.
A time-dependent fraction of this energy input is thermalized in the expanding ejecta, and thus the actual shape of the light curve is dictated by the competition between adiabatic degradation of internal energy into kinetic energy, and the loss of internal energy via radiation.

\begin{figure}
\includegraphics[width=0.48\textwidth]{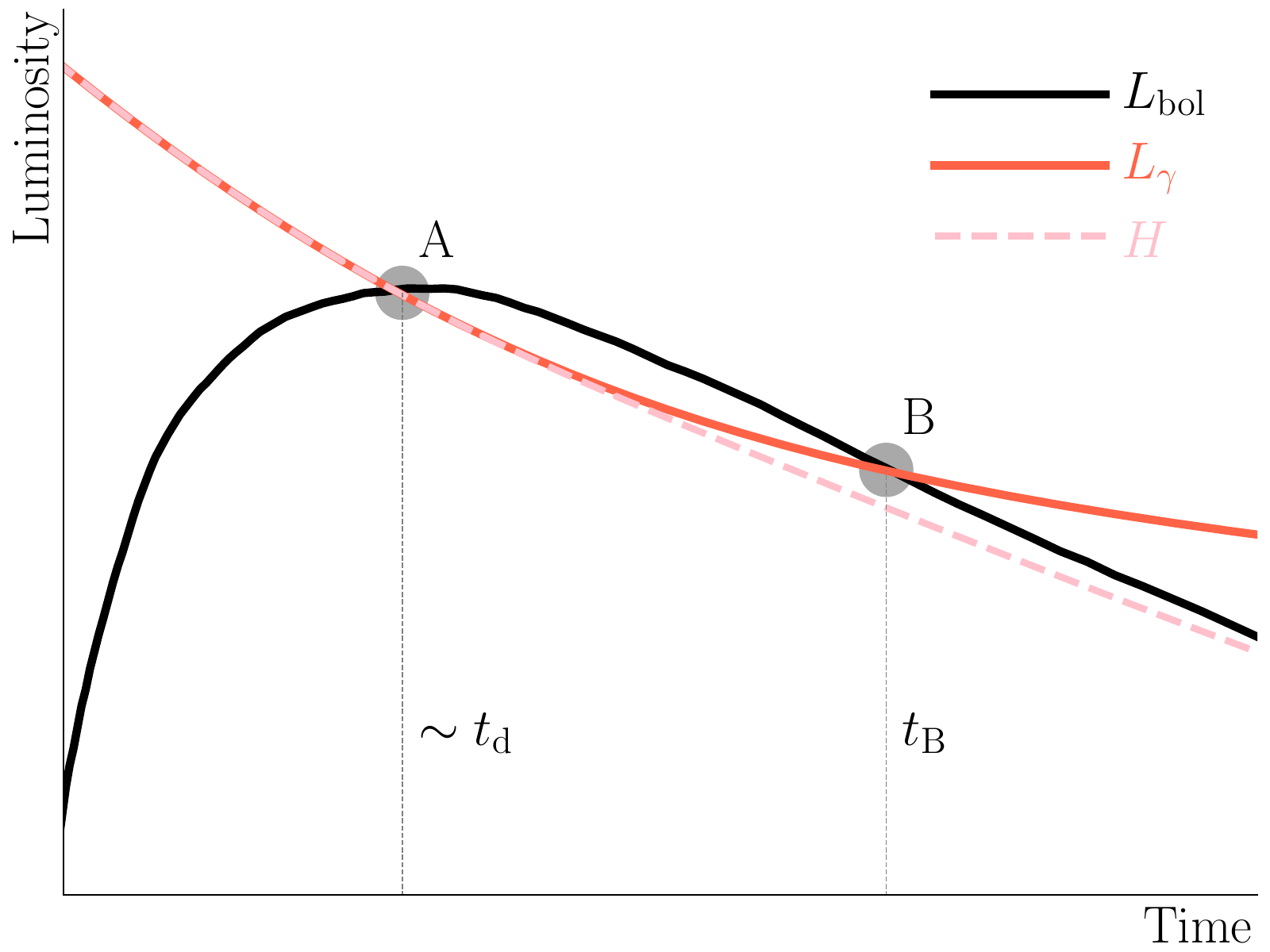}
\caption{Schematic representation of a Type-Ia light curve. The bolometric luminosity, $L_{\rm bol}$ 
         is shown as solid black, the instantaneous radioactive decay power, $L_\gamma$, is the red solid 
         curve, and the heating rate (thermalized fraction of $L_\gamma$), $H$, is shown as the dashed-pink curve. 
         The $L_{\rm bol}$ crosses $L_\gamma$ on two points - A and B. The former marks the well known 
         ``Arnett's'' rule, while this study aims to understand how the time of point B, 
         $t_{\rm B}$, is related to the time of point A, which is approximately the characteristic 
         diffusion timescale of the ejecta, $t_{\rm d}$. 
         \lFig{schem}}
\end{figure}

\Fig{schem} provides a schematic Type-Ia light curve. Since the explosion starts from a compact star, the ejecta begins its life as an opaque ball of plasma and thus the bolometric luminosity ($L_{\rm bol}$, black solid curve) rises at early times. Though the instantaneous radioactive power ($L_\gamma$, red solid curve) is very high at this time, the time that it takes for radiation to diffuse out of the ejecta is much larger than the age. As the expansion reduces the density, eventually the age of the ejecta surpasses the diffusion time of
\begin{equation}
t_{\rm d} = \Bigg[\frac{3\kappa M_{\rm ej}}{4\pi c v}\Bigg]^{1/2},
\lEq{diff}
\end{equation}
where $\kappa$, $M_{\rm ej}$, $c$, and $v$ are the opacity, ejecta mass, speed of light and ejecta velocity, respectively. Near this point, a significant amount of the deposited energy can be radiated rather than converted into kinetic energy for the first time, and the light curve reaches its peak. The well-known ``Arnett's rule'' \citep{Arn79,Arn82} defines this feature; he first showed that the light-curve peak must be close to the instantaneous decay power at that time (point A).

But this is not the only time the bolometric luminosity equals the decay power. Shortly after the peak of the light curve, there is a significant amount of radiation still trapped and diffusing outward in the ejecta \citep{Pin00a}. The bolometric luminosity remains greater than the decay power for some time while this ``excess'' energy is drained and the decay power falls onto the more slowly declining $^{56}$Co $\rightarrow ^{56}$Fe curve. Therefore the bolometric luminosity crosses the decay power again after the peak at point B.

Note that the bolometric luminosity crosses the decay power for a second time, not the actual heating rate ($H$, pink-dashed). Due to the escape of $\gamma$-rays, the non-thermalized fraction of the decay power increases with time, and so the heating rate crosses $L_{\rm bol}$ only once near the peak, and at late times it asymptotes to $L_{\rm bol}$. Also, the existence of point B may not be a universal feature of Type-Ia light curves. In extreme cases, the decay power may barely graze the bolometric luminosity at a single point near the peak, or it may not cross at all.

The aim of this study is to understand how the time, $t_{\rm B}$, of point B relates to the time, $\sim t_{\rm d}$, of point A. While the classical papers by \citet{Arn82} and \citet{Pin00a} provide semi-analytical model light curves, simple arguments are employed here to show why the time of point B is expected to be a constant multiple of the characteristic diffusion timescale of the ejecta. The claim is tested using both a set of Monte-Carlo radiation-transport calculations describing various possible ejecta configurations, and also observational data, to find that $t_{\rm B}/t_{\rm d}\approx1.7$. In the end, one of the main implications of this finding - the bolometric luminosity at 15 days after the peak is close to the radioactive decay power at that time, $L_{\rm bol}(t_{15})\approx L_\gamma(t_{15})$, is tested and I provide a discussion of how potentially this can be applied in the study of the observed width-luminosity relation \citep[WLR;][]{Phi99}.

\section{Semi-Analytical Arguments}
\lSect{args}

Energy conservation for the expanding ejecta implies that
\begin{equation}
\frac{{\rm d}E}{{\rm d}t}+P\frac{{\rm d}V}{{\rm d}t}+L_{\rm bol}=H,
\lEq{1st}
\end{equation}
where $E$, $P$, and $V$ are internal energy, pressure, and volume, respectively. The energy deposited from radioactive decay is stored, spent doing work on the expanding ejecta, and lost through the photosphere. Next, we employ two excellent and commonly invoked assumptions. First, the ejecta are homologously expanding in time with an isotropic velocity gradient, such that for a uniform density profile ${\rm d}V/{\rm d}t=4\pi v^3 t^2$. Second, the plasma is radiation-dominated, with $P=E/3V$. Under these assumptions,
\begin{equation}
\frac{{\rm d}E}{{\rm d}t}+P\frac{{\rm d}V}{{\rm d}t}=\frac{1}{t}\frac{{\rm d}(tE)}{{\rm d}t}.
\end{equation}
Writing the heating rate as $H=L_\gamma F_\gamma$, where $F_\gamma$ is the time-dependent deposition fraction, \Eq{1st} becomes
\begin{equation}
\frac{1}{t}\frac{{\rm d}(tE)}{{\rm d}t}=L_\gamma F_\gamma - L_{\rm bol}.
\lEq{main}
\end{equation}
This is equivalent to equation 10 of \citet{Kas10a} with a magnetar spin-down power $L_{\rm p}$ instead of radioactivity, and also to equation 2 of \citet{Kat13}, before time integration.

A critical piece in \Eq{main} is $F_\gamma$, where it must be a function that stays close to unity at early times, and then gradually asymptotes to zero at late times. First, consider early times, as they provide a simple way to derive the point A\footnote{The opacity changes with temperature and composition and may have significant time dependence \citep[e.g.,][]{Kho93}, which is ignored in this simplistic argument. For a more general derivation see \citet{Arn17}, and \citet{Kha18}.}. When $F_\gamma=1$, the condition $L_{\rm bol}=L_\gamma$ is satisfied only when ${\rm d}(tE)/{\rm d}t=0$. Employing the diffusion equation, one can derive an approximate relation,
\begin{equation}
L_{\rm bol}\approx\frac{4\pi Rc E}{3\kappa\rho V}=\frac{tE}{t_{\rm d}^2},
\lEq{tE}
\end{equation}
between bolometric luminosity and internal energy, where $R$ and $\rho$ are the radius and density. Taking time derivatives of both sides in \Eq{tE}, one sees that ${\rm d}(tE)/{\rm d}t=0$ is true when ${\rm d}L_{\rm bol}/{\rm d}t=0$, which is satisfied at the peak of the light curve (near the point A).

\begin{figure}
\includegraphics[width=0.48\textwidth]{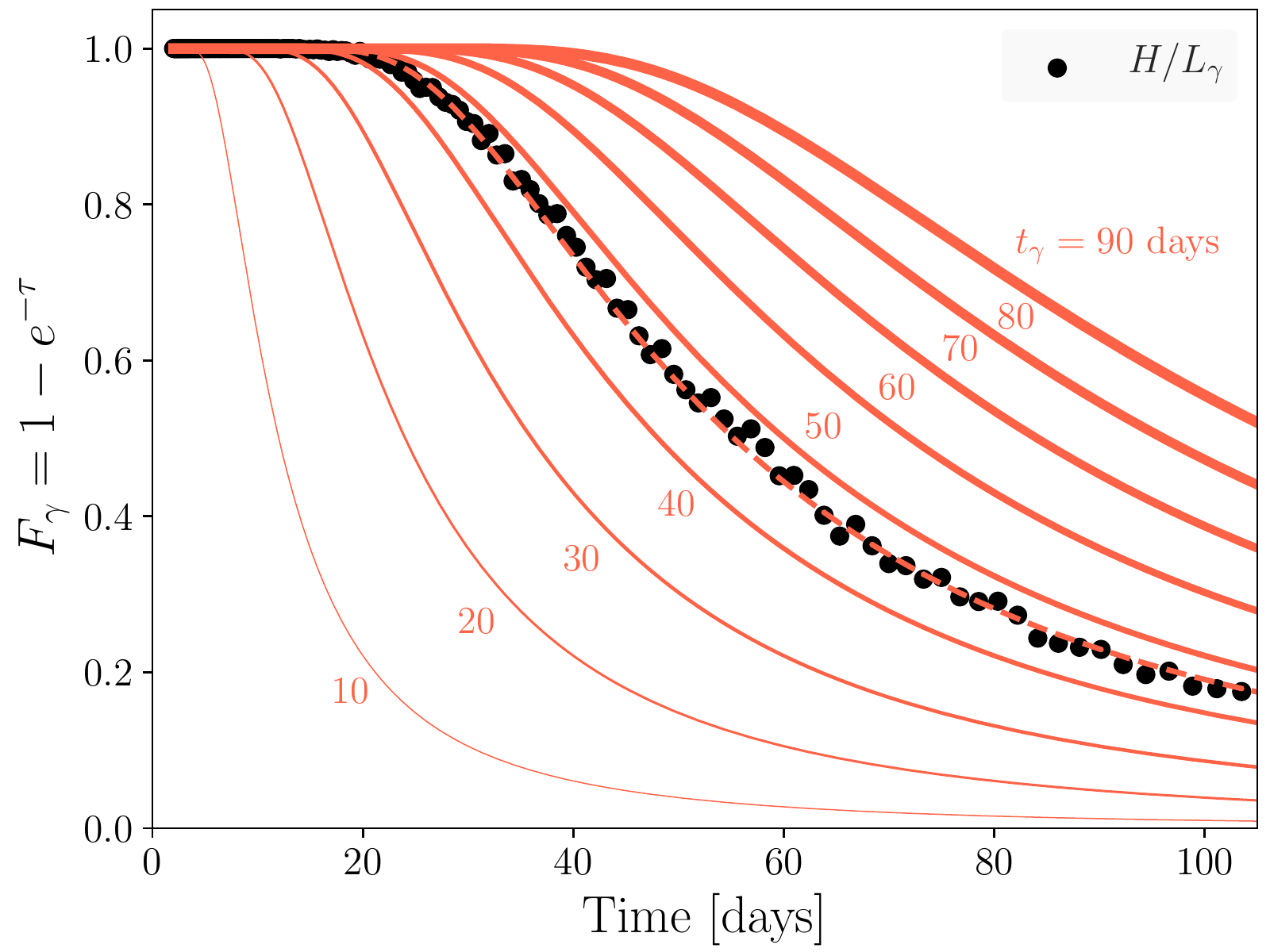}
\caption{Deposition fraction $F_\gamma=1-e^{-\tau}$ (red curves) as compared to 
         a sample radiation-transport calculation (black circles). The thicknesses of the 
         curves correspond to differing values of $t_\gamma$, where $\tau(t_\gamma)=1$. 
         Though it typically overpredicts the energy deposition at early times for small 
         $t_\gamma$, this prescription provides a good overall description of the 
         deposition fraction. The sample model is well fit with $t_\gamma=46$ days 
         (dashed red curve).
         \lFig{fgam}}
\end{figure}

As a more general and realistic description I adopt $F_\gamma=1-e^{-\tau}$ \citep[e.g.,][]{Pin01}. Here, $\tau=(t_\gamma/t)^2$ is the mean optical depth for gamma-rays, representing the mostly absorptive nature of Compton opacity. Due to the homologous expansion, the optical depth scales as $t^{-2}$ and the timescale $t_\gamma$ is chosen such that $\tau(t_\gamma)=1$. 

Positrons from the decay of $^{56}$Co will only start escaping the ejecta at very late times \citep[e.g.,][]{Mil99}, and therefore more formally $F_\gamma$ should apply to only gamma-component of $L_\gamma$. But for the sake of simplicity, the energetically less important positron contribution is ignored. A comparison of this prescription with a sample Monte-Carlo radiation-transport calculation (presented in \Sect{num}) is illustrated in \Fig{fgam}, where $t_\gamma\sim46$ days reproduces this specific model well.

With this description of $F_\gamma$, \Eq{main} becomes
\begin{equation}
L_\gamma e^{-\tau}+\frac{1}{t}\frac{{\rm d}(tE)}{{\rm d}t}=L_\gamma - L_{\rm bol},
\end{equation}
where $L_\gamma=L_{\rm bol}$ is satisfied on two conditions: (1) $L_\gamma e^{-\tau}={\rm d}(tE)/{\rm d}t/t=0$, equivalent to the condition for the point A discussed above for $F_\gamma\equiv1$; and (2)
\begin{equation}
\frac{{\rm d}(tE)}{{\rm d}t}=-L_\gamma e^{-\tau}t,
\lEq{scond}
\end{equation}
which is the condition for point B, and is satisfied only when $t=t_{\rm B}$. Unfortunately \Eq{scond} is not analytically integrable to elementary functions without approximations to the term $e^{t^{-2}}$. 

\begin{figure}
\includegraphics[width=0.48\textwidth]{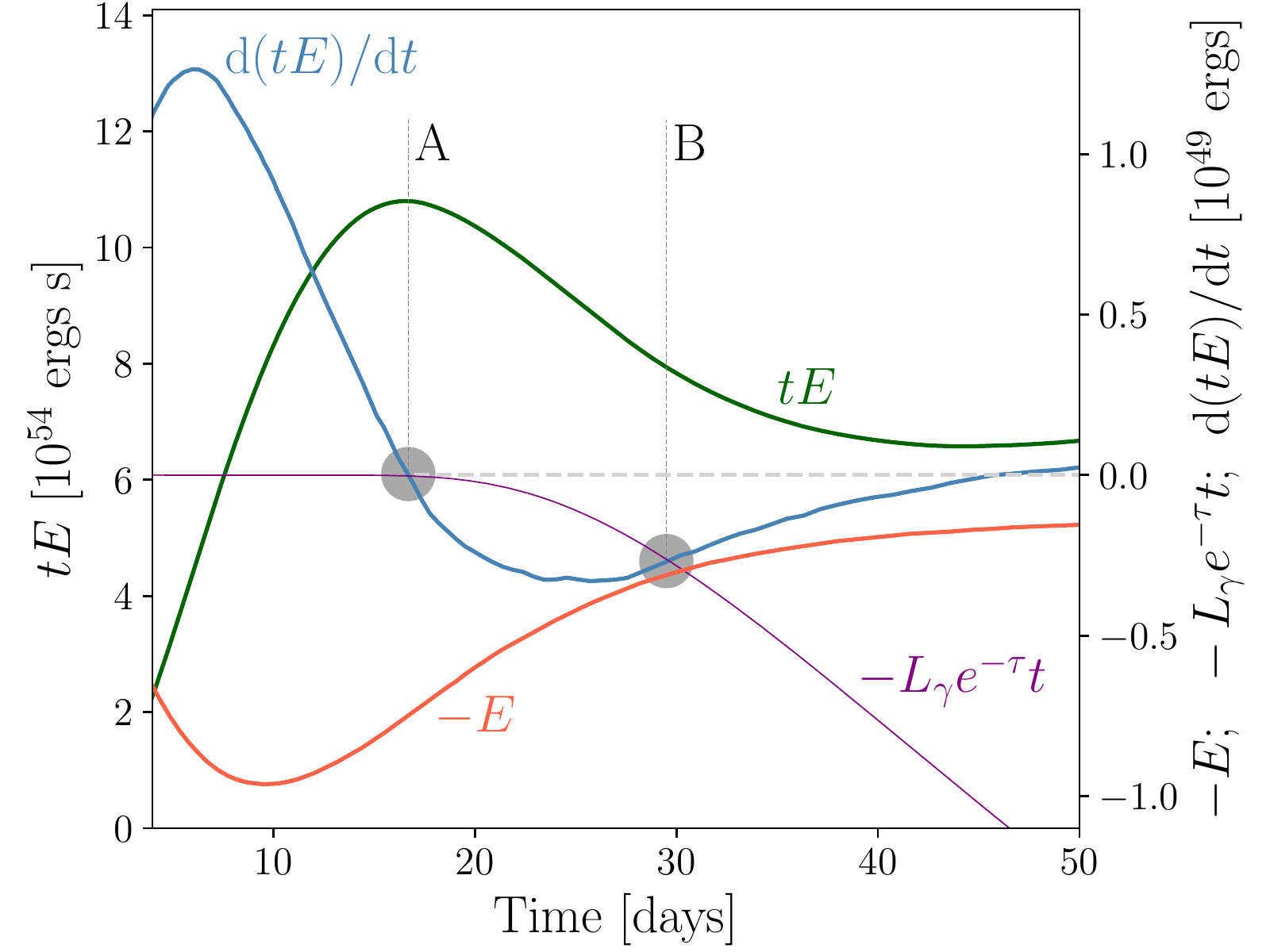}
\caption{The relevant quantities for \Eq{scond} are shown as a function of time for a sample 
         radiation-transport calculation. For clarity, the output from the 
         Monte-Carlo calculation is smoothed by a Savitzky-Golay filter. Note that the curves 
         for ${\rm d}(tE)/{\rm d}t$ and $-L_\gamma e^{-\tau}t$ first cross at point A near 
         zero (simplified condition for ``Arnett's rule''), and then cross again at 
         point B. Also, near the time of point B, ${\rm d}(tE)/{\rm d}t$ shares a common tangent 
         with $-E$ and therefore their derivatives are expected to be similar near this point.
         \lFig{econs}}
\end{figure}

But \Eq{scond} can be simplified using the fact that the internal energy changes in time roughly as ${\rm d}E/{\rm d}t\approx -2E/t$, near point B. This can be seen from the relationship between $E$ and ${\rm d}(tE)/{\rm d}t$, illustrated in \Fig{econs} for a sample model. Note the quantity ${\rm d}(tE)/{\rm d}t$ shares a common tangent with $-E$  near the time $t\sim t_{\rm B}$. Considering a general form of ${\rm d}E/{\rm d}t= xE/t$, where $x$ is a negative real number for $t>t_{\rm d}$, the time derivative of the left-hand term in \Eq{scond} becomes
\begin{equation}
\frac{{\rm d}}{{\rm d}t}\Bigg[\frac{{\rm d}(tE)}{{\rm d}t}\Bigg] = \frac{E}{t}(x^2+x).
\lEq{x}
\end{equation}
For example, the rate of change in the internal energy at the minimum of ${\rm d}(tE)/{\rm d}t$ can be found by solving $x^2+x=0$ for its non-zero root as ${\rm d}E/{\rm d}t=-E/t$. Since the curves ${\rm d}(tE)/{\rm d}t$ and $-E$ are tangents to each other near the point of interest, the time derivatives must also be similar at that point. Thus, the solution to $x^2+x=-x$ gives ${\rm d}E/{\rm d}t=-2E/t$ near $t\sim t_{\rm B}$. While this argument does not explain why the two curves are expected to share a common tangent near $t\sim t_{\rm B}$, numerical calculations (\Sect{num}) demonstrate that it is a good assumption.

Taking ${\rm d}E/{\rm d}t\approx-2E/t$, and replacing $E$ with \Eq{tE}, the \Eq{scond} becomes
\begin{equation}
L_{\rm bol}(t_{\rm B})t_{\rm d}^2 t_{\rm B}^{-1} \approx L_\gamma(t_{\rm B}) e^{-(t_\gamma/t_{\rm B})^2}t_{\rm B}.
\end{equation}
Recall that the condition for the point B is $L_{\rm bol}(t_{\rm B}) = L_\gamma(t_{\rm B})$, and thus the time of point B satisfies
\begin{equation}
t_{\rm B}/t_{\rm d} \approx e^{(t_\gamma/t_{\rm B})^2/2}.
\lEq{final}
\end{equation}
This implies that $\emph{if}$ the time of point B, $t_{\rm B}$, is proportional or slowly varying with $t_\gamma$ across a wide range of models, then one should expect it to also be a constant multiple of diffusion timescale $t_{\rm d}$. Furthermore, since the energy deposition terms cancel, \Eq{final} may also be applicable to other types of central power sources, such as magnetar spin-down \citep{Woo10,Kas10a} and black hole accretion \citep{Dex13}.

\section{Numerical Models and Observations}
\lSect{tests}

In this section, the arguments presented in \Sect{args} are explored with simple numerical calculations and with a set of bolometric measurements available from the literature.

\subsection{Radiation-transport Calculations}
\lSect{num}

The time-dependent Monte-Carlo radiation-transport code developed by \citet{Luc05} is used. The code includes Compton-scattering and photoelectric absorption for $\gamma$-ray transport, and employs gray transport for the optical radiation. Despite its simplicity, the resulting bolometric light curves are in good agreement with more advanced tools \citep[e.g., \SEDONA,][]{Kas06}.

\begin{figure}
\includegraphics[width=0.48\textwidth]{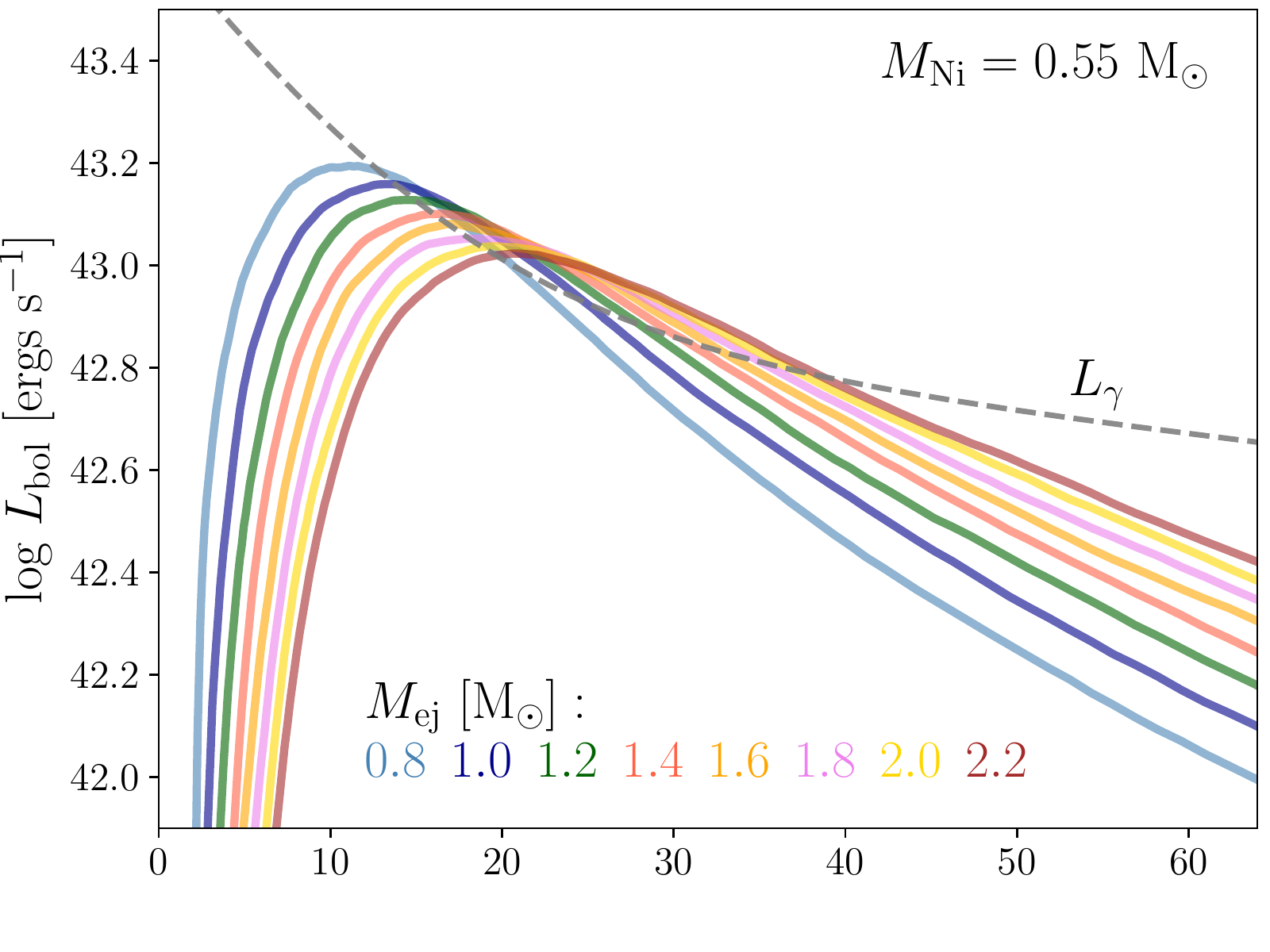}
\includegraphics[width=0.48\textwidth]{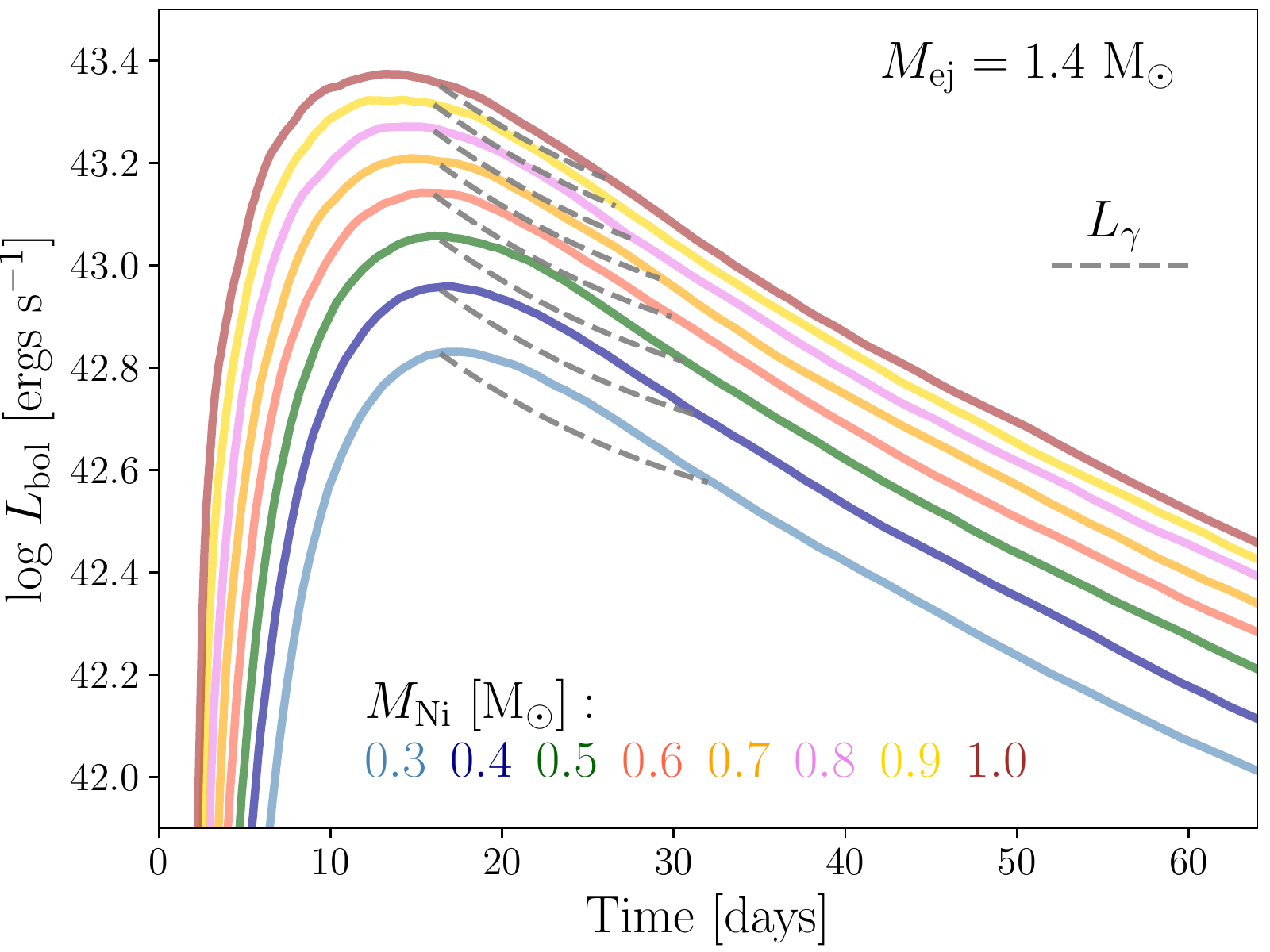}
\caption{Sample bolometric light curves from the grid of models computed with the Monte-Carlo 
radiation-transport code from \citet{Luc05}. The trends seen from the variation in ejecta mass (top) and 
         $^{56}$Ni mass (bottom) are consistent with prior calculations. The radioactive decay curve for 
         $M_{\rm Ni}=0.55$ \Msun\ is shown as a dashed gray curve in the top panel, while the decay curves 
         in the lower panel are shown only during the period when $L_\gamma < L_{\rm bol}$ (i.e. between the 
         points A and B). Note that this duration decreases as the ratio of $M_{\rm Ni}/M_{\rm ej}$ increases. 
         \lFig{lcs}}
\end{figure}

A small grid of 48 model light curves was generated based on a simple template. The ejecta are assumed to have a uniform density and the $^{56}$Ni mass is confined to the innermost region. The ejecta mass is varied between $0.8<M_{\rm ej}<2.2$ \Msun, and the $^{56}$Ni mass was varied between $0.2<M_{\rm Ni}/M_{\rm ej}<0.7$, so as to sample the wide range of possibilities emerging from various progenitor and explosion scenarios. For simplicity, a constant gray opacity of $0.1\ \rm cm^2\ g^{-1}$ is employed, and the initial outer radius and velocity of the ejecta were kept constant at $10^8$ cm and $10^{9}$ cm s$^{-1}$, respectively, for each model. Each spherically symmetric model is computed with 100 spatial zones and $5\times10^6$ radioactive matter packets. This choice is computationally cheap yet produces results with an acceptable level of noise. The final smooth light curves were obtained by applying a Savitzky-Golay filter with a second order polynomial.

Generic features of the resulting light curves are presented in \Fig{lcs} for $M_{\rm ej}=1.4$ \Msun\ with varying $^{56}$Ni mass, and for $M_{\rm Ni}=0.55$ \Msun\ with varying ejecta mass. With increasing ejecta mass, the diffusion timescale is longer, so the light curve evolves more slowly, reaching lower peak luminosities. With increasing $^{56}$Ni mass, the light curves are also broader, but they reach higher peak luminosities. These well-known general attributes are in good agreement with the results from many prior studies \citep[e.g.,][]{Pin00a,Woo07}. Note that, with increasing $M_{\rm Ni}/M_{\rm ej}$, the point A crossing is delayed with respect to the peak of the light curve, and $L_\gamma$ spends less time under $L_{\rm bol}$ (lower $t_{\rm B}/t_{\rm d}$). For extreme ratios, roughly when $M_{\rm Ni}/M_{\rm ej}\gtrsim0.8$ in these models, $L_\gamma$ never crosses $L_{\rm bol}$.

\begin{figure}
\includegraphics[width=0.48\textwidth]{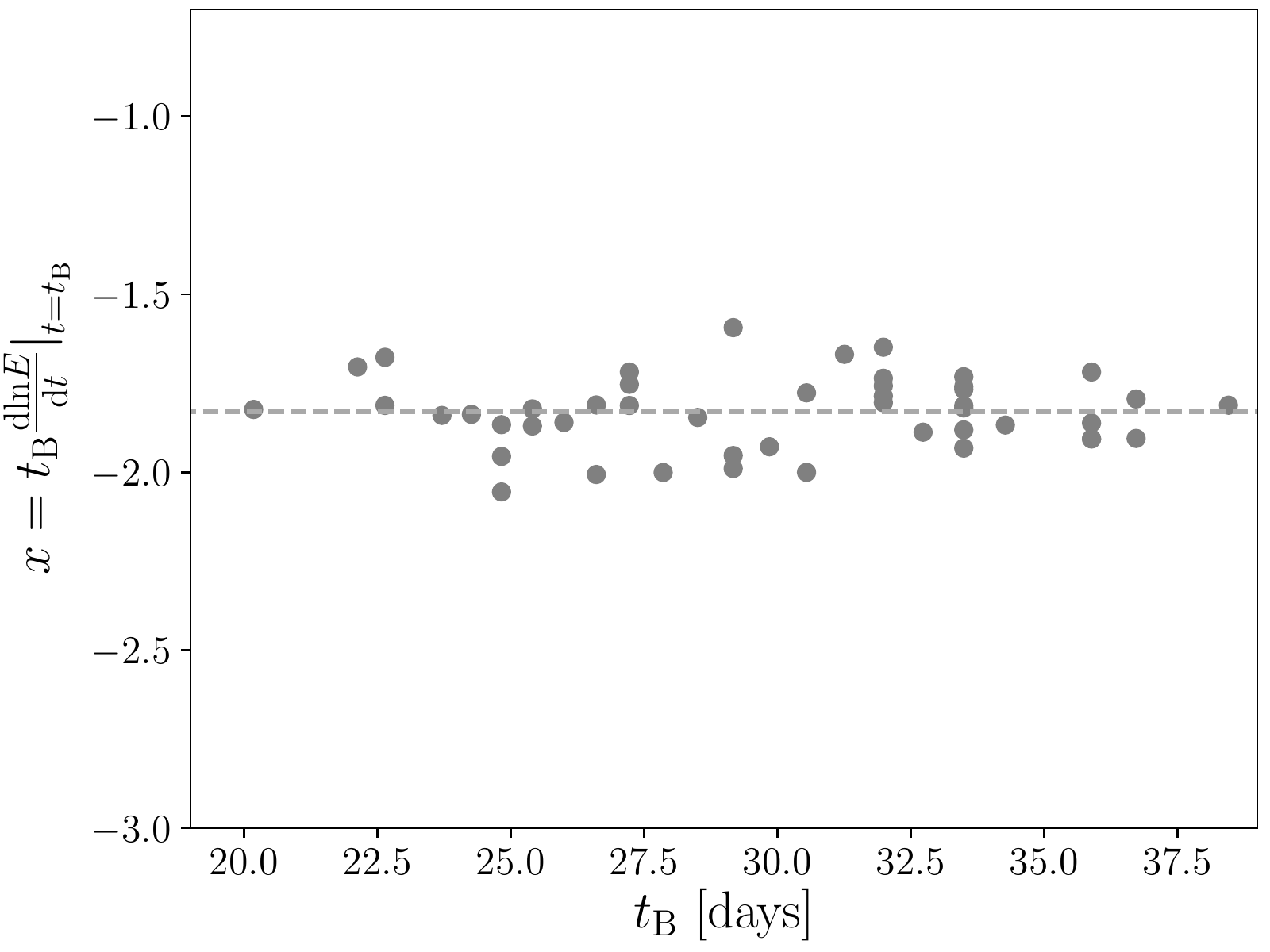}
\caption{Dimensionless quantity $x$ defined in \Eq{x} for all models at the time of point B.
         Note that it has a nearly constant value for a wide range of $t_{\rm B}$, 
         justifying the assumption of ${\rm d}E/{\rm d}t\approx-2E/t$ made 
         in deriving \Eq{final}.
         \lFig{xcheck}}
\end{figure}

Using this grid of models, we can numerically justify one of the key assumptions used in the derivation of \Sect{args} -- that the change in time-weighted internal energy is always tangent to the negative of the internal energy near point B, so that ${\rm d}E/{\rm d}t|_{t=t_{\rm B}}\approx-2E/t_{\rm B}$. \Fig{xcheck} illustrates the dimensionless quantity $x$ defined in \Eq{x}, evaluated at $t_{\rm B}$ for all models. It has a value close to $\sim -1.8$ for wide range of $t_{\rm B}$, justifying the assumption. 

The time $t_\gamma$, defined as $\tau(t_\gamma)=1$, is evaluated by fitting the models with the adopted functional form of $F_\gamma$, as illustrated in \Fig{fgam}. Finally, the time ratios in \Eq{final} are shown in \Fig{ratio} for all of our models. The apparent anti-correlation is not captured in the simplified arguments of \Sect{args}. But as can be seen in \Fig{lcs}, in models with increasing $M_{\rm Ni}/M_{\rm ej}$, the decay power spends less time under the bolometric luminosity, which results in a shorter $t_{\rm B}$. Also, some of the spread in the ratios of $t_\gamma/t_{\rm B}$ and $t_{\rm B}/t_{\rm d}$ are due to the assumptions of constant gray opacity and ejecta velocity, which results in a diffusion timescale that is not sensitive to $^{56}$Ni mass. Nonetheless, the points cluster within a fairly confined region, with only a 30\% variation in $t_{\rm B}/t_{\rm d}$, despite the very wide range of ejecta parameters. The ratio $t_\gamma/t_{\rm B}$ also clusters in a narrow range, as expected from \Eq{final}. Taking central values of $t_{\rm B}/t_{\rm d}=1.7$ and $t_\gamma/t_{\rm B}=1.6$, we see that the ratio of $t_\gamma/t_{\rm d}\approx2.7$ is also approximately constant.

As a simple sensitivity test, a subset of four Chandrasekhar mass models with varying $^{56}$Ni masses were recomputed with a broken power-law density profile, instead of the uniform one. Following the setup from \citet{Kas10b}, I adopt an inner segment of density, $\rho_{\rm i} \propto r^{-\delta}$ for
$v < v_{\rm t}$, and an outer segment, $\rho_{\rm o}\propto r^{-n}$ for $v\geq v_{\rm t}$, with $\delta=1$ and $n=10$. Here, $r$ is radius and $v_{\rm t}$ is a transitional velocity point connecting the two profile segments \citep[Equation (1) of ][]{Kas10b}. These power-law models have higher interior (innermost solar mass) and lower exterior density for a given kinetic energy and mass. The effect of higher central
density is dominant, and generally this leads to longer diffusion timescales and slower evolving light curves. However, the dimensionless quantity $x$ is still close to $-2$ in all four models, which implies $-E$ and ${\rm d}(tE)/{\rm d}t$ still share a common tangent near point B. The ratios of the timescales are systematically shifted by about $\sim 0.1$, but otherwise they show the same clustered pattern. While this simple test demonstrates the generic nature of these results, it should be noted that the post-peak ``hump'' seen in bolometric light curves is not modeled here. As noted by many prior studies \citep[e.g.,][]{She18}, the complex redistribution due to florescence requires non-gray ratiative transport without the assumption of local thermodynamic equilibrium.

\subsection{Comparison with Observations}
\lSect{obs}

Approximate bolometric light curves of actual SNe Ia supernovae can provide an independent, practical check on whether the ratios $t_{\rm B}/t_{\rm d}$ and $t_\gamma/t_{\rm B}$ are really constant across a wide range of explosions. Unfortunately, bolometric measurements are difficult and are not frequently published. In this work, a small but modern set of spectroscopically \emph{normal} Type-Ia light curves compiled by the Nearby Supernova Factory project\footnote{https://snfactory.lbl.gov} is employed. For details on the sample selection and construction of the bolometric luminosity see \citet{Sca14},\citet{Chi13} and \citet{Ald02}.

The reconstructed $M_{\rm Ni}$, $M_{\rm ej}$ values, and ``joint'' host galaxy reddening values from Tables 2 and 3 of \citet{Sca14} are adopted. The radioactive decay power is assumed to cross the bolometric curve near the time of peak, which is estimated through the characteristic diffusion timescale with corresponding $M_{\rm ej}$, but with constant $\kappa=0.1\ \rm cm^2\ g^{-1}$ and $v=10^9\ \rm cm\ s^{-1}$. From the original list of 19 events, 3 cases, \texttt{SNF20080717-000}, \texttt{SNF20080913-031}, and \texttt{SNF20080918-004}, were discarded due to their irregular luminosity evolution such that $L_\gamma$ never crosses $L_{\rm bol}$, or crosses it more than twice.

\begin{figure}
\includegraphics[width=0.48\textwidth]{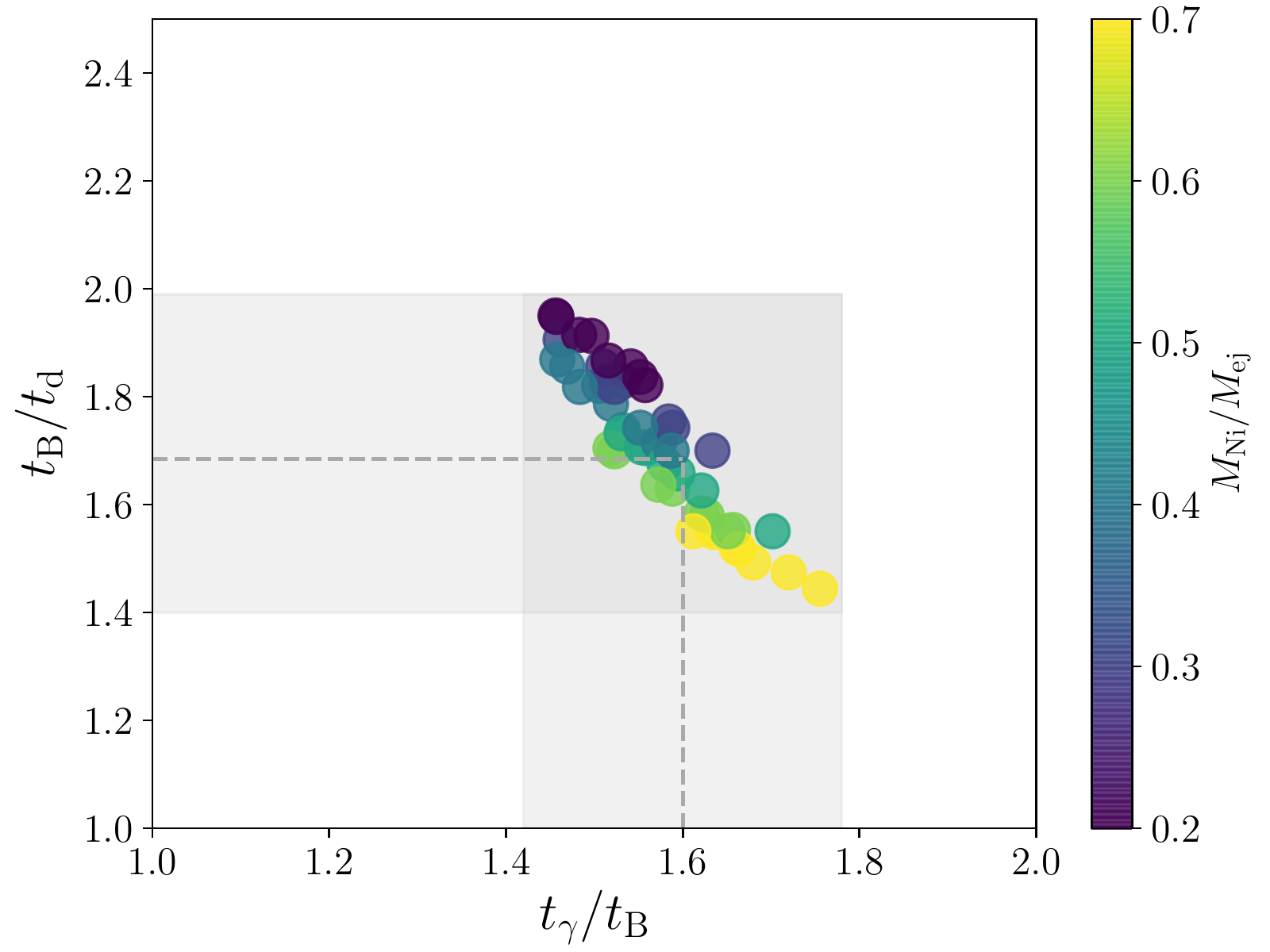}
\caption{Time ratios derived in \Eq{final} for all 48 radiation-transport calculations.
         Despite the models spanning a very wide range of parameter space, the resulting ratio of the  
         point B time and the diffusion timescale varies by only $\sim30$\%, around 
         $t_{\rm B}/t_{\rm d}\approx1.7$. The ratio $t_\gamma/t_{\rm B}$ is also confined to a 
         narrow range around $\sim1.6$, as expected from \Eq{final}.
         \lFig{ratio}}
\end{figure}

\Fig{obs} shows the ratio $t_{\rm B}/t_{\rm d}$ measured for the remaining 16 events. In all cases, the ratio is distributed in a narrow range around a central value of $t_{\rm B}/t_{\rm d}\approx1.65$. This is in a close agreement with the simple radiation-transport calculations presented in \Fig{ratio}. Note that these 16 events had a relatively narrow range in ejecta mass $0.9<M_{\rm ej}<1.4$ \Msun, but $0.3<M_{\rm Ni}/M_{\rm ej}<0.65$, similar to the range covered by the model light curves. Therefore the tighter distribution of $t_{\rm B}/t_{\rm d}$ as compared to the models may be indicating a systematic difference between the reconstructed parameters from this observational sample and the simple radiation-transport calculations used in \Sect{num}.

Recently \citet{Wyg17} proposed a novel method of estimating $t_\gamma$ from the bolometric light curve, and on the same sample of observations \citep{Sca14}, they have found that it spans a narrow range between 30 and 45 days. The values of $t_{\rm B}$ measured for these events in this study span between 19 and 30 days. This implies $30<t_\gamma<48$ days for $t_\gamma/t_{\rm B}=1.6$ (\Fig{ratio}), and is in excellent agreement with the range measured in \citet{Wyg17}. This agreement suggests that $t_\gamma$ can be estimated without integration on bolometric measurements; instead a simple estimate of the peak time (i.e. in B-band) is used, $t_{\rm d}\approx t_{\rm peak}$ and thus $t_\gamma=2.7t_{\rm d}$.

\section{Luminosity 15 Days after Peak}
\lSect{15d}

Simple analytic arguments, numerical calculations, and observational tests indicate that $t_{\rm B}\approx 1.7 t_{\rm d}$ for wide range of Type-Ia explosion light curves. But how can this relation can be useful?

Consider the following aspects of Type-Ia light curves. (1) Both theoretical and observational studies have shown that the simple version of ``Arnett's rule'' works with an accuracy of about $10$\% in Type-Ia explosions \citep[e.g.,][]{Blo13,Str06} - in most cases $L_\gamma$ crosses $L_{\rm bol}$ near the peak of the light curve at $t\sim t_{\rm d}$. (2) The range of characteristic diffusion timescale, or the range of the observed light-curve rise time, is not that great. It is almost always between 10-25 days, which, according to $t_{\rm B}\approx 1.7 t_{\rm d}$, implies $t_{\rm B}$ will be roughly between 7 and 18 days \emph{after the peak}. (3) The rate of change ${\rm d}L_{\rm bol}/{\rm d}t$ and ${\rm d}L_\gamma/{\rm d}t$ are not too different near $t_{\rm B}$, unless the light curve is evolving too fast.

\begin{figure}
\includegraphics[width=0.48\textwidth]{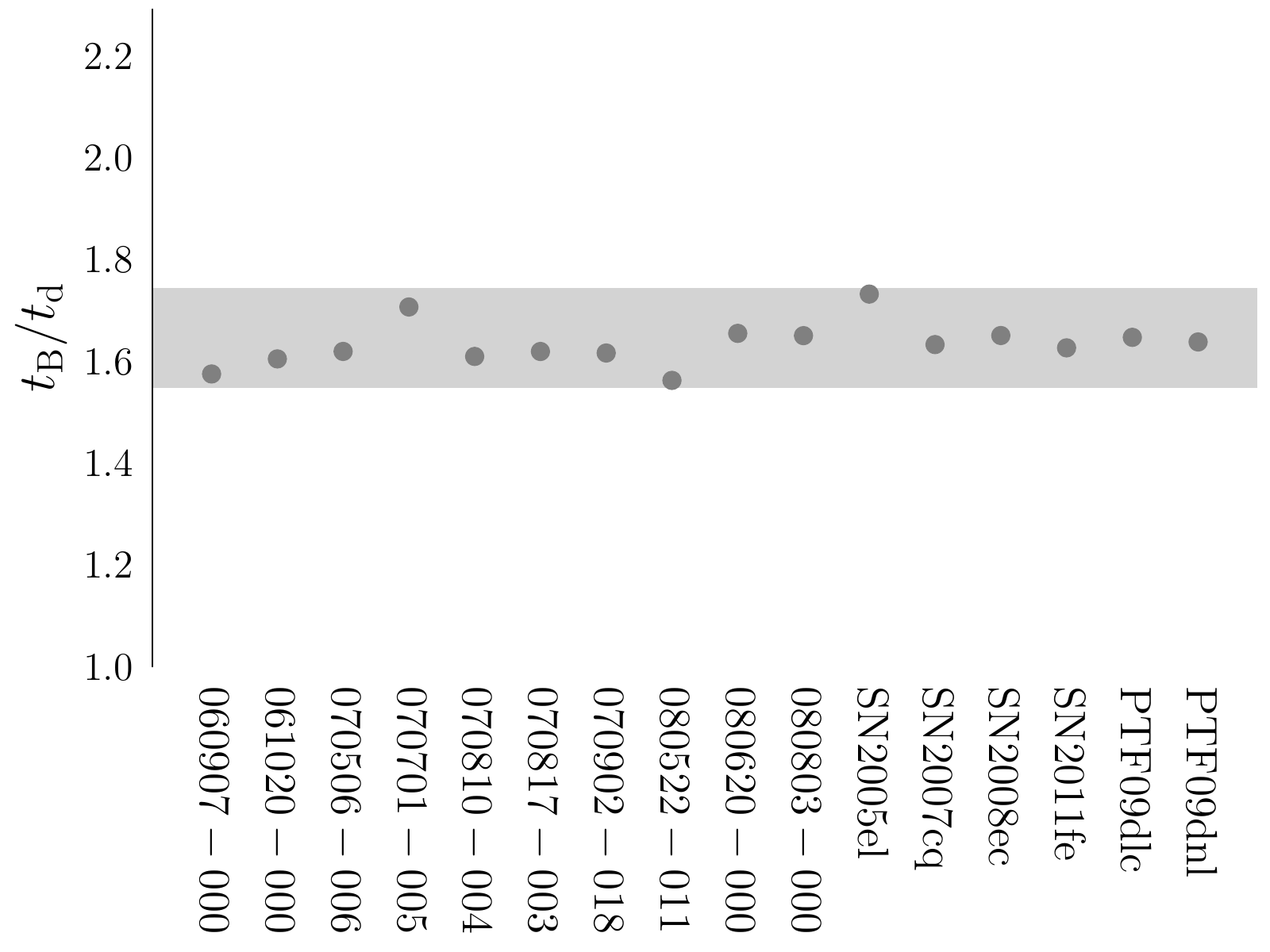}
\caption{The ratio $t_{\rm B}/t_{\rm d}$ as measured for the set of Type-Ia bolometric light 
         curves from \citet{Sca14}. The Neaby Supernova Factory event names have been 
         shortened to their discovery dates (i.e., \texttt{060907-000} is 
         \texttt{SNF20060907-000}). Three events with peculiar luminosity evolution have been 
         excluded from the analysis. For the 16 remaining events, the ratio falls in a narrow range 
         centered around $t_{\rm B}/t_{\rm d}\approx1.65$, consistent with the calculations 
         presented in \Fig{ratio}.
         \lFig{obs}}
\end{figure}

\begin{figure*}
\centering
\includegraphics[width=0.99\textwidth]{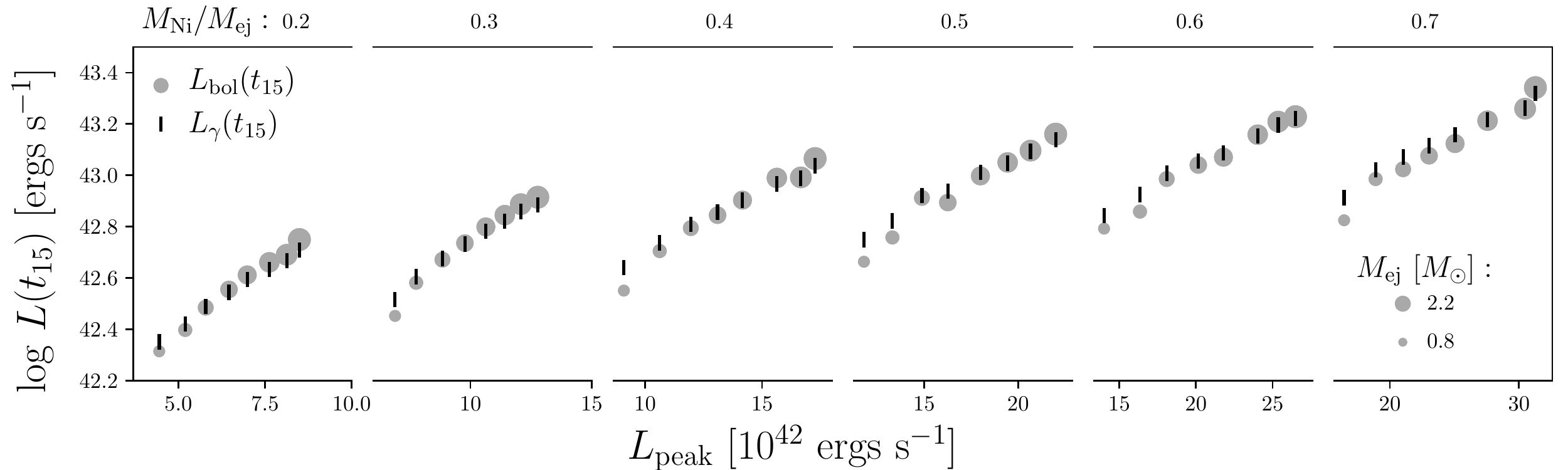}
\caption{The bolometric luminosity at 15 days after the peak (gray circles) in all 48 model light curves 
         presented in \Sect{num} is compared to the radioactive decay power at that time (black bars). 
         For a given ratio of $M_{\rm Ni}/M_{\rm ej}$, the ``true'' luminosity is overpredicted at 
         low ejecta mass (small circles) and underpredicted at high ejecta mass (larger circles). 
         In almost all of these models the luminosity at 15 days after the peak is predicted to 
         better than 10\%.
         \lFig{LL}}
\end{figure*}

\begin{figure*}
\includegraphics[width=0.5\textwidth]{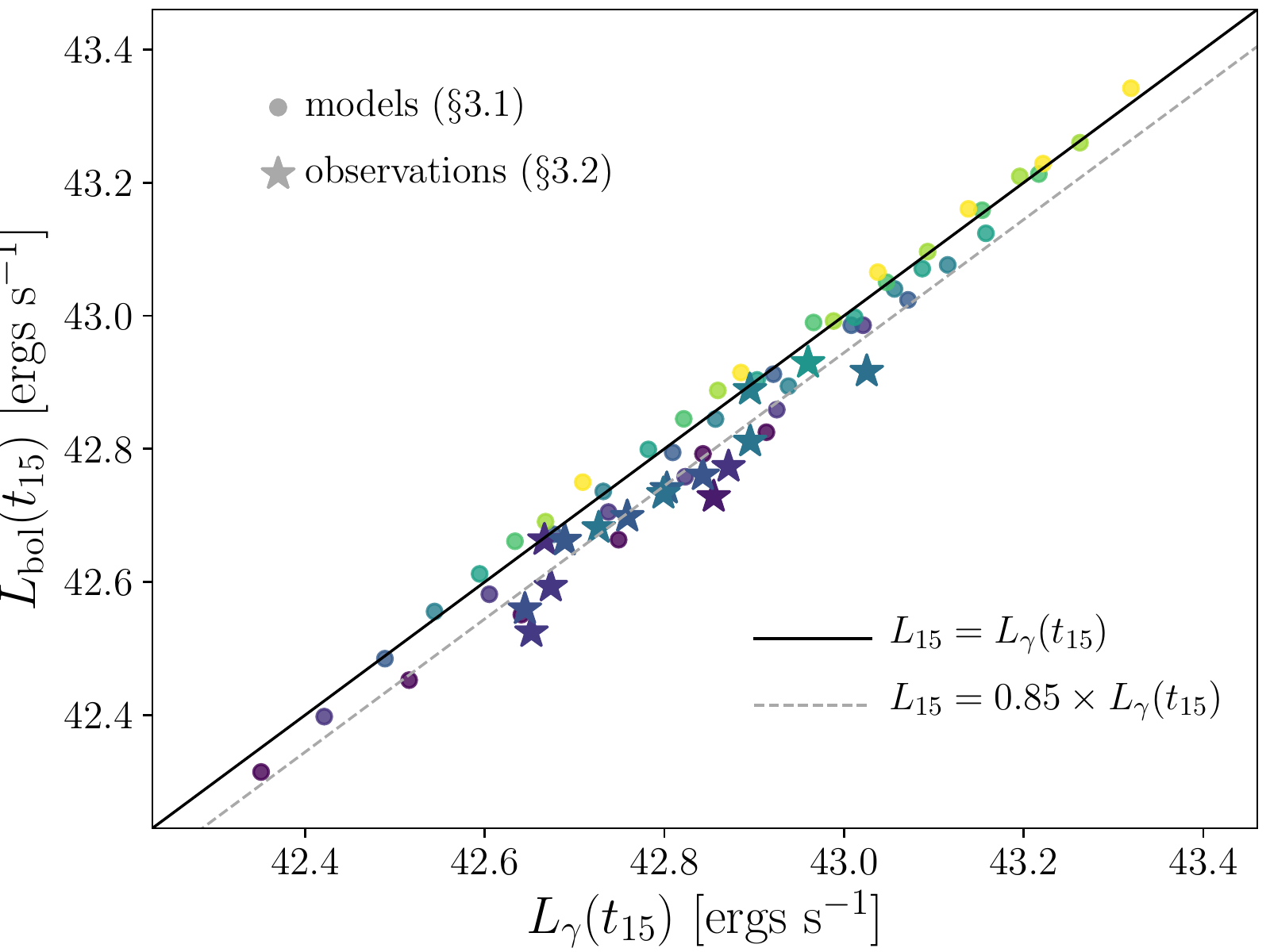}
\includegraphics[width=0.5\textwidth]{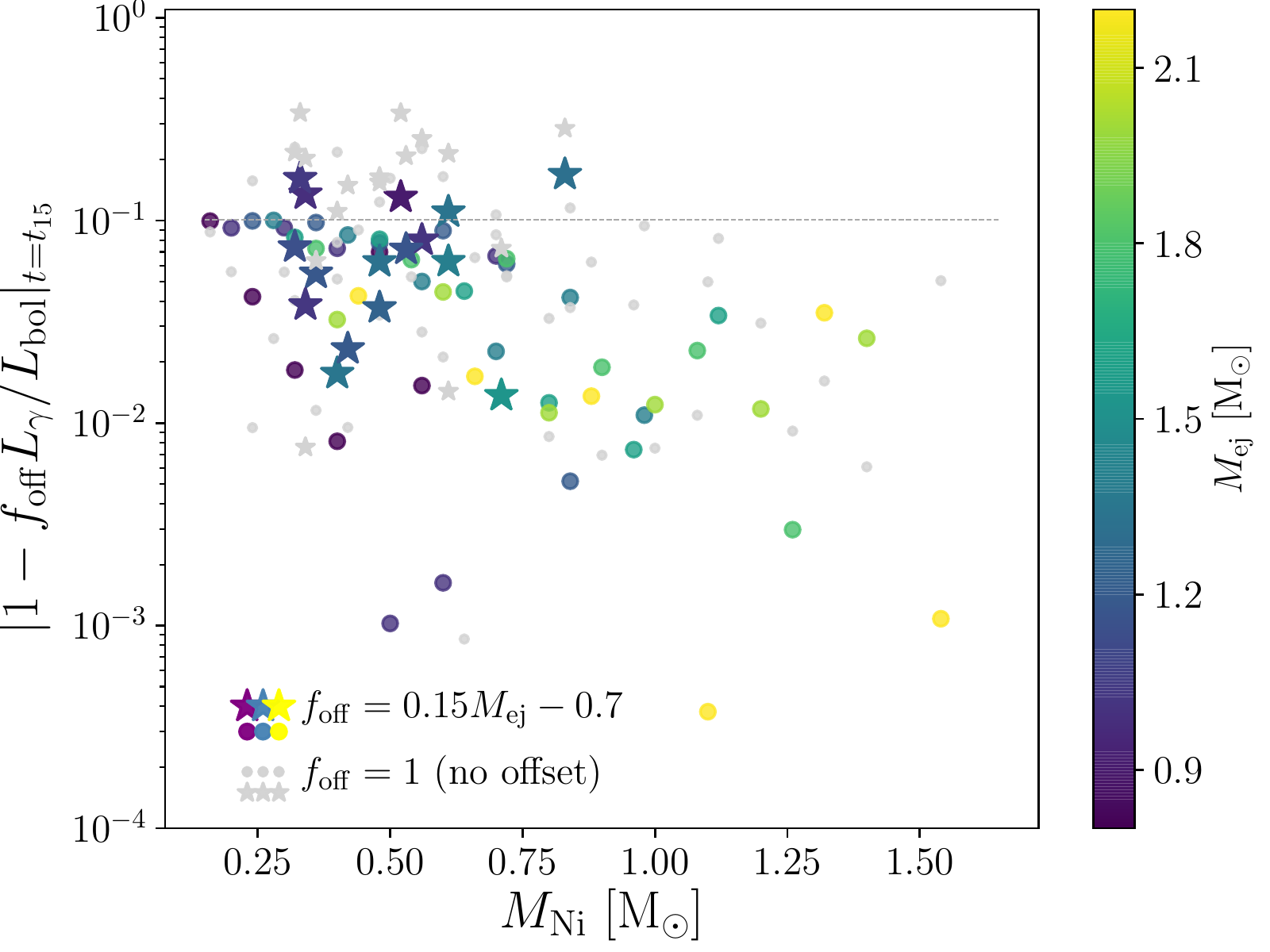}
\caption{(Left) The correlation between $L_{15}$ and $L_\gamma(t_{15})$ for all models (circles) 
         and observational events presented in \Sect{obs} (stars). All model values lie close to 
         the $L_{15}=L_\gamma(t_{15})$ line (black solid, $\lesssim$10\% deviation), while the observational 
         values are systematically overpredicted by $L_\gamma(t_{15})$. The overall fit is improved,  
         with a small offset of $f_{\rm off}=0.85$ (dashed gray).
         (Right) The absolute fractional difference between $f_{\rm off}L_\gamma(t_{15})$ and 
         $L_{\rm bol}(t_{15})$, for an ejecta mass dependent offset of $f_{\rm off}=0.15 M_{\rm ej}-0.7$ (colored circles and 
         stars) and without offset $f_{\rm off}=1$ (gray circles and stars). 
         \lFig{perc}}
\end{figure*}

These aspects suggest that the time $t_{\rm B}$ will typically happen a few days earlier than the time 15 days after the peak, $t_{15}$, and therefore normally $L_\gamma(t_{\rm B})>L_{\rm bol}(t_{15})$. However, because $L_\gamma$ is a decreasing function of time, the decay power at $t_{15}$ should be much closer to the bolometric luminosity at that time, i.e. the following is true no matter $t_{\rm B}>t_{15}$ or $t_{\rm B}< t_{15}$: 
\begin{equation}
|L_\gamma(t_{15})-L_{\rm bol}(t_{15})|<|L_\gamma(t_{15})-L_{\rm bol}(t_{\rm B})|
\end{equation}
In order to test how close $L_\gamma(t_{15})$ is to $L_{\rm bol}(t_{15})$, the luminosities measured at $t_{15}$ are shown as a function of peak luminosity, $L_{\rm peak}$, in \Fig{LL}. The ``true'' values at $t_{15}$ have been measured from the model light curves computed in \Sect{tests} (gray circles), and compared to the corresponding instantaneous radioactive power at that time (black bars). 

In general, the light curves that reach higher peak luminosities will have higher luminosities at $t_{15}$, and for a given ratio of $M_{\rm Ni}/M_{\rm ej}$, this trend correlates with the ejecta mass (gray circle sizes). Note that $L_\gamma(t_{15})$ always overpredicts the luminosity for the lowest ejecta masses, and it underpredicts it for the highest ejecta masses. This reflects the fact that $t_{\rm B}>t_{15}$ for lower ejecta masses and the opposite at higher ejecta masses. Overall, excluding the handful of rapidly evolving models with the lowest ejecta and $^{56}$Ni masses, the radioactive decay power at $t_{15}$ closely matches the bolometric luminosity at that time (to within $\lesssim10$\%).

The left panel of \Fig{perc} shows the correlation between $L_{15}$ and $L_\gamma(t_{15})$. All of the radiation-transport calculation values (circles) fall closely around the one-to-one correlation. Also shown are the values measured from the sample of 16 SNe Ia supernovae presented in \Sect{obs} (stars). As mentioned earlier, the simple calculations from \Sect{num} do not model the post-peak ``hump'' due to florescence, and therefore the observational values are slightly less luminous for a given $L_\gamma(t_{15})$. Since most of the poorly behaving cases are due to overpredicted luminosities, the overall fit can be improved by introducing a small offset factor, $L_{15}=f_{\rm off}L_\gamma(t_{15})$, with $f_{\rm off}=0.85$ (dashed gray).

Note that most of the overpredicted cases are of lower ejecta mass, and thus the overall fit can be further improved with a simple ejecta mass dependent offset, $f_{\rm off}=0.15M_{\rm ej}-0.7$. The right panel of \Fig{perc} illustrates the absolute fractional difference between $f_{\rm off}L_\gamma(t_{15})$ and $L_{\rm bol}(t_{15})$. Without any correction factor ($f_{\rm off}=1$), nearly all model values (gray circles) are below 10\%, but the observed values (gray stars) are off by as much as 35\%. With mass-dependent offset, all of the models and majority of observed events lie below 10\% (colored circles and stars).

Given the simplistic nature of the numerical models and the uncertainties in parameters estimated from the observations, it is hard to determine which comparison deserves more weight as a test. But it is encouraging to see that $L_\gamma(t_{15})$ very closely matches with $L_{15}$ from a set of models that span a wide range of parameter space, and that a simple offset can greatly improve the overall agreement.

\section{Conclusion}
\lSect{conclude}

This study investigates the properties of SN Ia supernova light curves through energy conservation arguments, radiation-transport calculations, and observational tests. The main finding is that for a wide range of parameters, the time when the instantaneous radioactive decay power crosses the bolometric luminosity for the second time, after the peak of light curve, appears to be a constant multiple of the characteristic diffusion timescale of the ejecta. For these sets of simulations and observed SNe Ia, this constant turns out to be $\sim1.7$, i.e.,
\begin{equation}
t_{\rm B}\approx1.7t_{\rm d}.
\end{equation}
It has been shown that the gamma-ray escape timescale is also related with the diffusion timescale roughly as, $t_\gamma \approx 2.7 t_{\rm d}$.

The primary implication of this finding is that this relation suggests the bolometric luminosity 15 days after the peak must be very close to the instantaneous radioactive decay luminosity at that time, i.e.,
\begin{equation}
L_{\rm bol}(t_{15})\approx L_\gamma (t_{15}).
\end{equation}
It may serve as a simple to tool that connects the observables of the WLR to a physical description of the ejecta.

A calibrated version of this relation that works for individual band-absolute magnitudes is needed for a more practical application on WLR. For instance, the $B$-band magnitudes will evolve significantly faster than the bolometric magnitude after the peak, so without any calibration, it is likely that the 
$B$-band magnitude 15 days after the peak will be systematically overpredicted in all cases. These types of effects may be explored by employing more advanced radiation-transport tools, e.g., \texttt{SEDONA}, \citep{Kas06}, \texttt{STELLA}, \citep{Bli06}, \texttt{CMFGEN}, \citep{Hil12}) and \texttt{JEKYLL}, \citep{Erg18}, to see if reliable calibrations can be built for specific bands. A proper radiative transfer treatment is crucial in modeling the post-peak ``hump'' in bolometric light curves as well.
  
There may also be other subtle uses of this relation, where it could be employed in the interpretation of certain poorly sampled light-curve measurements. For instance, if the peak of the light curve is missed, but the $^{56}$Ni mass is estimated from the nebular spectra, then $t_{\rm B}$ can be measured from the light curve. This relation implies $t_{\rm d}\approx t_{\rm B}/1.7$ and the peak luminosity would be $L_{\rm peak}\approx L_\gamma(t_{\rm d})$.
 
As was emphasized originally in \citet{Arn82}, the Type-Ia explosion light curves are physically simpler than core-collapse Type-Ib/c, where there is a much weaker association between the main heating agent and the kinetic energy source. However, since the energy source terms cancel in the derivation of \Eq{final}, the proposed relation will also approximately hold for Type-Ib/c explosions. It would be an interesting future project to explore this relation in Type-Ib/c light curves, including the cases that are dominantly powered by energy sources other than radioactivity.

In general, a larger sample of observations with good constraints on the explosion date, host reddenning and preferably with independently measured $^{56}$Ni masses \citep[e.g. nebular spectra, ][]{Chi15}, would go a long way in demonstrating the usefulness of this relation.

\section*{Acknowledgments}
I wish to thank the anonymous referee for providing a thorough review of this work. I also thank Todd Thompson, Stan Woosley, Anthony Piro, Chris Kochanek, Gantumur Tsogtgerel and Maximilian Stritzinger for many helpful discussions. The radiation-transport code used in this work was developed by Leon Lucy, who sadly passed away earlier this year. I wish to thank him for his many pioneering contributions to astrophysics. Support for this work was provided by NASA through the NASA Hubble Fellowship grant \#60065868 awarded by the Space Telescope Science Institute, which is operated by the Association of Universities for Research in Astronomy, Inc., for NASA, under contract NAS5-26555.

\Leg{Software:} \texttt{matplotlib} \citep{Hun07}, \texttt{numpy} \citep{Van11}.

\end{document}